# Design and implementation of tools to build an ontology of Security Requirements for Internet of Medical Things

Daniel Naro, Jaime Delgado, Silvia Llorente, Amanda Palomo
*Departament d'Arquitectura de Computadors (DAC),*
*Universitat Politècnica de Catalunya · UPC BarcelonaTECH,*
*C/ Jordi Girona, 1-3, 08034 Barcelona, Spain*
*E-mail: daniel.naro@upc.edu*

**Abstract**. When developing devices, architectures and services for the Internet of Medical Things (IoMT) world, manufacturers or integrators must be aware of the security requirements expressed by both laws and specifications. To provide tools guiding through these requirements and to assure a third party of the correct compliance, an ontology charting the relevant laws and specifications (for the European context) is very useful. We here address the development of this ontology. Due to the very high number and size of the considered specification documents, we have put in place a methodology and tools to simplify the transition from natural text to an ontology. The first step is a manual highlighting of relevant concepts in the corpus, then a manual translation to XML/XSD is operated. We have developed a tool allowing us to convert this semi-structured data into an ontology. Because the different specifications use similar but different wording, our approach favors the creation of similar instances in the ontology. To improve the ontology simplification through instance merging, we consider the use of LLMs. The responses of the LLMs are compared against our manually defined correct responses. The quality of the responses of the automated system does not prove to be good enough to be trusted blindly, and should only be used as a starting point for a manual correction.

## 1. Introduction

Security requirements present in standards and specifications play a crucial role when evaluating security and safety of medical systems. The emergence of new medical devices, some of them based in the Internet of Things (IoT), opens new threats in the development of medical systems. Those devices are not always controlled by medical institutions but by patients and, in the most challenging case, they may be even at patients' home, out of the sight (and surveillance) of system administrators.

On the other hand, the number of standards and specifications considering security requirements is increasing and regulations are becoming stricter in order to protect patients' life and wellbeing when using these new devices, which may include from simple measurement devices, like thermometers, to actuators like insulin pumps.

With the aim of facilitating regulations accomplishment, we are developing a security requirements ontology in the context of the MedSecurance [1] European research project. It is based on existing standards and specifications that some partners of the project need to support the systems they are integrating in the project, which include both software and physical devices.

Internal of Medical Things (IoMT) is a special case of IoT which, as the name indicates, focuses on the medical world. The paper first introduces the IoMT environment, and then our need to identify all its security and safety coverable aspects. To do this, we use a corpus of documents, relevant to the IoMT domain. Then, we present how the knowledge contained in the different laws (or regulations) and specifications (standards) is structured into an ontology, i.e., at transforming the currently unstructured natural text into structured knowledge by means of an ontology. To simplify this process, we both use existing tools, in particular Obsidian MD [2], and develop our own ones to go from an unstructured knowledge (contained in natural language text) to a semi-structured one (expressed as an XML document based on an XML Schema defined by us) and finally to a structured one (expressed as an ontology). Afterwards, we show the result of using our tools in the case of IoMT. This exercise highlights some of the challenges we have faced during the development, discussing some of them. For example, we talk about unused knowledge contained in the corpus, or on the usage of Large Language Models (LLM) [3] to find possible improvements on our generated ontology through its simplification. Finally, we draw some conclusions and future work.

## 2. Background

### 2.1. Analysis of the State of the art related to security and IoT

Our search for the state of the art is focused on ontologies providing information regarding the security landscape in machine-to-machine communication and IoT. A main point of interest is that the ontology should be based on a corpus of specifications. Only one publication [4] integrates specification requirements, but we were not able to retrieve their ontology. This means that our main need is not covered by current publications. Other papers help in the selection of security mechanisms [5][6], study the information security [4][7] or the communication channel [8]. Finally, one paper focuses on the assessment of the security [9]. Furthermore, concepts such as safety are not addressed because these publications are oriented towards the IoT world in general and not specifically to IoMT. This also has implications in regard to the provided data classification.

**Table 1.** Analysis of the state of the art

|        | Ontology is available | Includes standards' requirements | Includes theoretical attacks | Includes real world attacks | Includes control measures | Includes data classification | Includes safety concepts |
|--------|:---:|:---:|:---:|:---:|:---:|:---:|:---:|
| [5][6] | ✓ | ✗ | ✓ | ✓ | ✓ | ✓ | ✗ |
| [4]    | ✗ | ✓ | ✗ | ✗ | ✗ | ✗ | ✗ |
| [8]    | ✓ | ✗ | ✓ | ✓ | ✓ | ✗ | ✗ |
| [9]    | ✓ | ✗ | ✓ | ✗ | ✓ | ✗ | ✗ |
| [7]    | ✓ | ✗ | ✗ | ✗ | ✓ | ✓ | ✗ |

The publication landscape, summarized in Table 1, has motivated us to construct a new ontology ensuring the exhaustive modelling of the requirements and concepts appearing in their formulation, which also takes into account safety.

### 2.2. Internet of Medical Things (IoMT)

The technologies developed for the Internet of Things (IoT) are also relevant to the medical world, becoming the Internet of Medical Things (IoMT). The promise is to be able to allow remote tracking of, for example, patients' constants, and, potentially, also administering treatments remotely (for example, through an insulin pump).

This usage opens new challenges: the generated and communicated data has implications regarding both security and safety. There are already specifications on how to protect private data, for example ISO 27001 [10]. Other specifications are being written which address the specificities of IoMT, for example the P2933 Standard for Clinical Internet of Things Data and Device Interoperability with Trust, Identity, Privacy, Protection, Safety, Security (TIPPSS) [11]. Furthermore, IoMT is also affected by laws, as described by Medical Device Coordination Group (MDCG) both in MDCG 2019-11 [12] and MDCG 2019-16 Rev.1 [13].

The development of IoMT devices and services thus requires to be aware of the concepts considered in the specifications and laws, and the requirements stated therein. For example, the manufacturer will be required to document certain pieces of information in documents addressed to different actors. The manufacturer will also be required to understand which risks of the system might lead to outcomes affecting safety, and, finally, understand what kind of severity class the risk belongs to (i.e., to which legal or standardized severity group the risk belongs to, such as potentially lethal).

Specifications and laws usually define concepts that are not easily translated into technical terms. The development of tools for the guidance through these documents and automation of the IoMT system's evaluation would therefore be of great help both for manufacturers and final users.

### 2.3. MedSecurance

One of the aims of MedSecurance [1], a European research project, is to address the previously described problem. One of the project's outputs will be a set of tools to ease the evaluation of IoMT devices' security characteristics to be used by the provider (i.e., device manufacturer or service provider) and consumer (of device or service) to determine the possible risks, ensure their correct mitigation and generate the necessary information to assure the security and safety of the product.

Industrial and user partners involved in the project defined a set of requirements, which the resulting toolbox should address. These requirements include specifications and regulations which shall be supported: these were the starting point for the list we considered and which we introduce in clause 2.4.

Beyond the specifications obtained in the project and described in clause 2.4, other tools developed in the project must rely on the knowledge contained in relevant specifications: for example, understanding which are the actors at play in an IoMT ecosystem, or what are the different types of data (configuration, backup, personal data or medical data, among others). To this end, it is important to have an ontology which covers the knowledge contained in a corpus of specifications.

MedSecurance also aims at providing feedback to MDCG, describing specific measures to the guidance on cybersecurity based on the tools implemented and lessons learnt during the development of the project.

### 2.4. Specifications' coverage

The creation of an ontology focusing on security and safety requirements in the IoMT domain is not only a cornerstone for developing guidance tools, but also to give feedback to the authors of the existing specifications. As an example, we will be able to give feedback to the European guidance documents on cybersecurity [12][13], described by the Medical Device Coordination Group (MDCG). A possible feedback item is reporting on a relevant aspect not being covered by this guidance document.

To help us discover such blind spots, we first need a representation of the full picture. To construct this representation, we intend to rely on as many specifications identified as relevant to IoMT as possible. Through listing all requirements stated in each specification, we can construct a clearer picture of the relevant set. As introduced in clause 2.3, partners defined a list of specifications and regulations important for their use case. This list was further extended to obtain the final collection of documents transcribed in the ontology and itemized next:

1. ISO 11073 [14][15], cybersecurity for Medical / health device communication standards. It is focused on which elements of a medical device could be a vulnerability, and which types of attack could be carried out on it. Requirements defined in this standard are limited to the identification of risk in the development phase
2. ISO 14971 [16], risk management for medical devices. It is focused on the analysis of the risks and how to reduce their impact during the development of the medical device
3. ISO 27001 [10], information security management. It mainly manages the information security within an organization, but it does not specifically focus on medical devices
4. ISO 80001 [17], risk management for medical IT-network. It standardizes the risk management related to a medical IT-network: i.e., a network which needs to take into account security and safety for the medical devices and health software connected to it. This standard has several parts, including one that establishes a security case framework. The objective is to simplify the communication between a Medical Device Manufacturer and Health Delivery Organization so that the Health Delivery Organization has the information required to support the risk management of the IT networks
5. ISO 82304 [18], safety and security of health device software. It provides requirements for the safety and security of health software products that are software-only products
6. MDCG 2019-11 [12], guidance on qualification and classification of software for medical device. It is

focused on guiding throw the European law regarding medical device software classification, and partly the technical consequences of such a classification
7. MDCG 2019-16 [13], guidance on European's law mandates on cybersecurity for medical device. It integrates concepts from ISO 27001 and ISO 14971
8. IEEE P2933 [11], Clinical Internet of Things (IoT) Data and Device Interoperability with TIPPSS - Trust, Identity, Privacy, Protection, Safety and Security. The goal is to establish the framework for the TIPSS principles and device validation and interoperability for healthcare, including wearable IoT devices, hospital devices and other connected healthcare systems. It is applicable to devices including hardware, firmware and software
9. SESIP [19], Security Evaluation Standard for IoT Platforms (SESIP). It defines a methodology to reduce cost, complexity and effort of security evaluation and certification. It focuses on the requirements to be covered in order to create a secure platform, providing the mechanism to deploy applications, their updates and identify instances
10. MEHARI [20], Methode Harmonisee d'Analyse de Risques. It is a French risk assessment methodology that aims to help organizations identify and analyze risks to their information systems and develop appropriate risk management strategies
11. SABSA [21], Sherwood Applied Business Security Architecture. It is a risk management framework that provides a structured approach to designing and implementing security solutions in an organization. It is based on the concept of enterprise security architecture, which means that it integrates business, information technology, and security architecture to align the security strategy with the organization's business objectives

## 3. Design of the toolkit

To address the needs of the MedSecurance project, we have started from a non-structured input (each specification, i.e., natural language document), from which we have extracted the information, structured it and then transformed it into an ontology. To further refine the content, the pipeline also proposes LLM-enabled tools which further transform the structured information or ontology to improve its content through simplification (merging synonyms or closely related concepts). These steps are detailed in the next clauses, but they can be summarized as follows:
1. Text highlighting and content linking: the natural language content is annotated using ObsidianMD. This program allows us to create a collection of interconnected nodes.
2. Early searchable information: the collection of interconnected nodes can be queried as a graph. We use the PageRank algorithm to identify the central concepts from our set of notes.
3. Translation of corpus into XML+XSD: we manually structure the information of interest into a pair of XML document and its schema. This step defines the structure of information which will appear in the ontology.
4. Automatic translation to OWL/RDF: we use the tools we have developed to automatically translate the content represented in the XML+XSD pair into an ontology.
5. Considering improvements through LLMs: the generated ontology might present rough edges which could be improved automatically thanks to LLMs. In our case, we have multiples entries which are similar to one or more other entries. The use of LLMs can be used to attempt to automatically detect this mergeable pairs.

### 3.1. Text highlighting and content linking

As introduced in clause 2.3, we need to process a corpus of specifications covering as much as possible the IoMT domain.

The information contained in these specifications must be extracted and given some structure. As we do not have the golden set to ascertain the quality of a program's extraction, we must perform this step by hand. To improve ergonomics of this first step, we have relied on Obsidian MD [2]. This program allows the creation of notes (i.e., files) in the Markdown format [22], which might be interlinked among themselves (as shown in Figure 1). For example, an Obsidian note containing the text "user of device" might link both to the notes named "user" and "device".

To extract the knowledge from the specifications using Obsidian, we copy natural language content into Obsidian notes, recreating the clause hierarchies using folders. These notes leverage the labelling features of Obsidian, by integrating a tag specifying that the content belongs to a clause from one of the specifications. Then, the relevant keywords are highlighted by creating links to (still) empty notes.

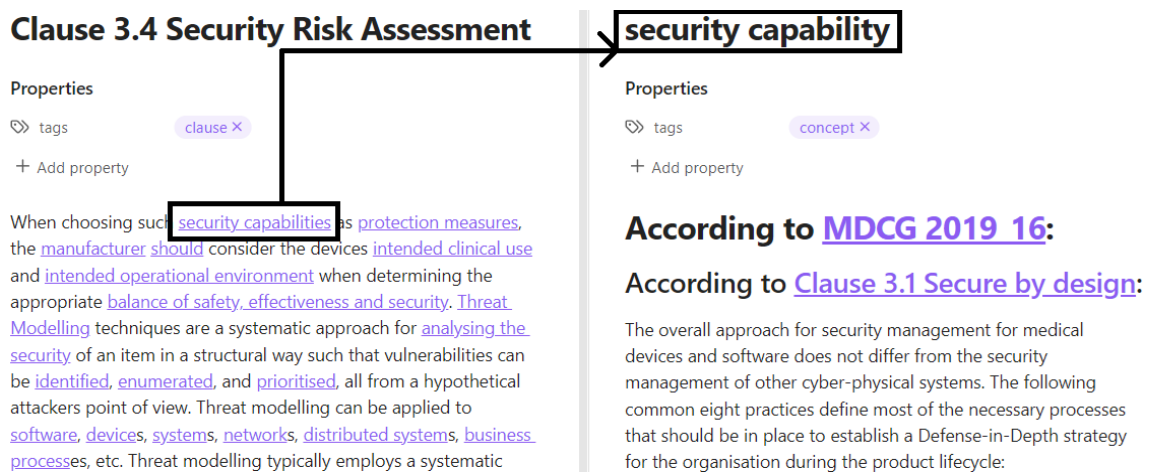

**Figure 1** Obsidian's interlinkage of notes

It is up to the user to decide how many nodes are needed. As previously introduced, there are wordings such as "user of device" which can be deconstructed into multiple levels. Some elements such as "data breach detected event" can be deconstructed into three levels of their own type:
- Event: "data breach detected event"
- Action: "data breach detection"
- Attack: "data breach"

When annotating the content, we have tried to maximize the number of annotated levels, in order to cover as many foreseeable uses as possible.

As previously mentioned, Obsidian's added value is the interlinkage of notes. As in our methodology the notes of concepts appearing in the standards are created empty, we only have clause notes (i.e. text copy from the standards) pointing to empty concept notes. This would seemingly defeat the purpose of the exercise due to the lack of links between concept notes. To remedy this, without introducing a greater workload, we have developed a tool which iterates over those notes tagged as clause and populates the linked notes with the text of the paragraphs where the concept appeared (prepending it with a heading indicating the origin specification and corresponding clause). By doing so, we ensure that the concepts notes are interconnected, without going into each note and adding its content.

### 3.2. Early searchable information

Having Obsidian notes for the relevant concept interlinked with other relevant concepts appearing together in the specifications' text enables us to have a semi-structured dataset. The notes are still in natural language form, yet the linkage is machine readable, and can be represented as a graph. This allows mapping out which concepts are addressed by a given specification, and which concepts are closely related.

The graph nature can be readily leveraged to compute the relevance of the different concepts through the PageRank algorithm [23]. This method gives an idea of the most relevant concepts in the corpus of specifications. However, the graph and the PageRank results do not provide the semantics of the linkage between concepts.

Similarly, travelling through the graph from a concept notes allows to retrieve the collection of relevant natural text clauses (heading and text). This can be of relevance to quickly discover which concepts are central to which standard, but also as a way to retrieve the input material for an LLM [3] query.

### 3.3. Translation of this corpus into XML + XSD

To improve the searchability on the data, we structure it by manually translating it to XML format alongside its XSD schema [24].

The XML can be found under https://dmag-upc.github.io/XMLXSDOntologyBuilder/midMedSecSecurity.xml or in the git repository https://github.com/DMAG-UPC/XMLXSDOntologyBuilder, which can be obtained with the command git (src/test/resources/generated_Testing_EntireRequirements.xml).

The XSD schema can be found under https://dmag-upc.github.io/XMLXSDOntologyBuilder/schemaMedSecSecurity.xsd or in the git repository https://github.com/DMAG-UPC/XMLXSDOntologyBuilder, which can be obtained with the command git (src/main/resources/input/RequirementsSchemas.xsd). The entry point to our translation are the "requirements": i.e., the root element of the XML resulting from the translation is a sequence of translated (and thus semi-structured) requirements.

We refer to requirements as the statements formulated in the different specifications saying that a certain actor may, should or shall do a specific action.

All encountered requirements share some basic characteristics, for example a requirement:
- has a name
- refers to an actor in charge of it
- may be related to an event
- might have to be complied with prior to the compliance of another requirement

Then, depending on the requirement, other semantic relationships are needed. We leverage the inheritance features provided by XSD, to define a basic requirement structure which is then extended in different child structures addressing the representation needs. At the moment, and to cover the requirements encountered in the current version of specifications' corpus, we have identified, among others, the need for the requirement definitions (_requeriment has been removed from the name of the requirement for clarity purposes) described in Table 2.

**Table 2**. Identified requirements

| Requirements currently identified | |
|---|---|
| AvoidingEventDesignRequirement | CommunicationRequirement |
| ConsiderationRequirement | DefenceAgainstRequirement |
| DocumentationRequirement | EnsuringRequirement |
| EstablishmentRequirement | EvaluationRequirement |
| FollowingRequirement | IdentificationRequirement |
| MaintenanceRequirement | ManagingRequirement |
| MitigationRequirement | MonitoringRequirement |
| PerformingRequirement | ProhibitedImplementationRequirement |
| ProhibitedPerformingRequirement | ValidationRequirement |

Each of these requirement types define the relevant relationships with other concepts: both in nature and semantics of the relationship. Figure 2 shows the schema for the ensuring_requirement element. It provides a choice of what needs to be ensured: either a given concept, or its property, the execution of an action or the compliance with a standard. The semantic of the relationship is established through the name of the element in the choice, while its nature is mandated by the nature of the defined XML element.

### 3.3.1 Decisions required when defining the XSD

As introduced in clause 3.3, we want to translate this schema into an ontology, which will then be populated with the contents of our XML file.

To achieve this, we had to introduce some rules for the definition of the XSD:

1. There shall be no nameless types. In XSD, an element can define its type in situ, without giving it a proper name. Nevertheless, as we need to define the ontology's classes with their names, we rely on the XSD type's name and mandate that it is always provided.
2. XSD choices shall be annotated with a name. When creating the ontology's Data or Object Property, we require a name for this relationship. As XSD's choices are nameless, we mandate the provisioning of an annotation to provide their name.
3. The root element shall be a list of elements of interest. The transformation program expects to find within the root element a sequence of elements to translate into ontology instances. Each element in the hierarchy is then traversed to discover every instance to be created in the ontology.

```xml
<xs:complexType name="EnsuringRequirement">
    <xs:complexContent>
        <xs:extension base="ns:Requirement">
            <xs:choice>
                <xs:annotation>
                    <xs:appinfo>EnsuringRequirement_choice</xs:appinfo>
                </xs:annotation>
                <xs:element name="ensured_concept" type="ns:Content"/>
                <xs:element name="ensured_action" type="ns:action"/>
                <xs:element name="compliance_with_standard" type="ns:Standard"/>
                <xs:element name="ensured_property" type="ns:ContentProperty"/>
            </xs:choice>
        </xs:extension>
    </xs:complexContent>
</xs:complexType>
```

**Figure 2.** Schema for EnsuringRequirement type

### *3.3.2    Downside: Duplication of nodes*

The reliance on XML elements has positive effects upon the manual task of structuring the extracted information (such as being able to rely on existing tools to provide auto-completion regarding the structure of the XML), yet it also introduces some challenges.

The main challenge is the duplication of certain elements across the XML document. For example, many requirements refer to the "manufacturer". Therefore, there are at least as many elements of type "actor" having the name "manufacturer". Other duplicated concepts can have more child elements. For example, there will be multiple instances of "benefit-risk tradeoff analysis", which in every case is decomposed into:

- an "analysis" named "benefit-risk tradeoff analysis"
- of a "tradeoff" named "benefit-risk tradeoff"
- between a first concept named "benefit"
- and a "risk" named "risk"

The transformation tool, described in clause 3.4, will thus have to identify the presence of an already known element, and, instead of translating it into a new ontology instance, reference the previously defined one.

### **3.4. Automatic translation to OWL/RDF**

For the representation of the ontology we choose the pair OWL/RDF as RDF [25] provides an ontology serialization, and OWL [26] provides mechanism to represent complex relationships between concepts. The automatic translation of the XML and XSD files into an ontology combining OWL and RDF requires two steps. First, the XSD schema is translated into an ontology containing only classes and afterwards the XML file is read and used to populate the ontology with instances. The resulting ontology is available under https://dmag-upc.github.io/XMLXSDOntologyBuilder/ontoMedSecSecurity.owl or in the git repository https://github.com/DMAG-UPC/XMLXSDOntologyBuilder, which can be obtained with the command git (src/test/resources/expected_Testing_RequirementSchema.xml).

For the first step, we do not rely on any existing library. Instead, we have used the JAXB package [27] to automatically translate the XSD schemas defining RDF and OWL into Java classes. The resulting Java code only required minimal changes to ensure the proper behavior regarding the mix of XML namespaces or collisions in the elements' naming. We have used the ontologies serialized by Protégé [28] as the target output for our tool, both for the classes' definition and the definition of their properties.

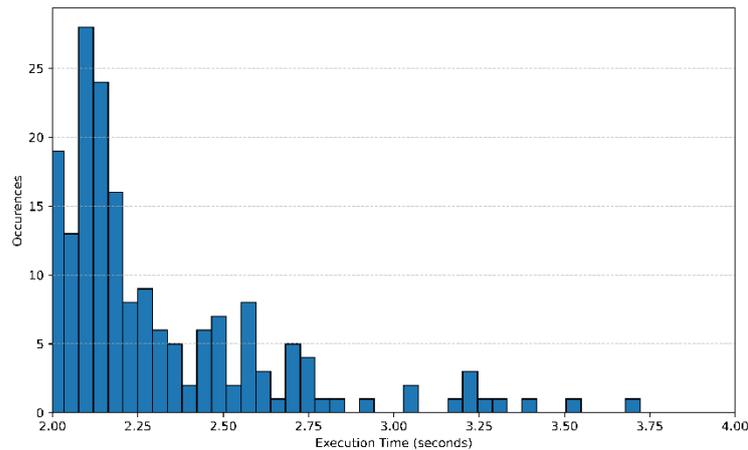

**Figure 3**. Histogram of execution times

The result of this first translation step is an automatically generated ontology equal to what we would have obtained by manually generating the classes and properties. For the current size of our corpus (1403 requirements, more than 20k concepts), the translation takes an average of 2.36s on a machine with an i5-1240P and 16GB of RAM. The histogram of processing times is shown in Figure 3, showing a higher count close to the average, and a tail which corresponds to the initial execution, where the automatic local optimization of the code might not yet be in place.

The second step is then to populate the ontology with the instances. To do that, we iterate over the XML file, and generate as many required instances in the ontology as needed.

When reading solely the XML file, we do not have any information on the exact nature of a given element. For example, the "actor" in charge of a requirement, the "recipient" of a communication, and the "recipient" of a document are in all cases instances of the "actor" class. To resolve this, we rely on the Xerces library [29], which allows us to provide both the XML and XSD, thus enabling us to discover the correct ontology class for a given XML-element. This approach also means that the XSD is discovered at execution time and not at compilation time, as would be the case when using JAXB. This greatly simplifies the modification of the XML/XSD, as no considerations are to be given to the translation besides the ones explained in clause 3.3.1.

The actual creation of instances of classes and properties is performed using the OWLAPI library [30]. In clause 3.3.2, we introduced the challenge of duplicated elements. Our automatic translation tool ensures that, if a given element has already been included in the ontology, then we rather refer to the existing instance. This determination is performed by analyzing the attributes and elements of the element and its descendant. The tool's source code is available in the git repository https://github.com/DMAG-UPC/XMLXSDOntologyBuilder.

With this step, we have finally transformed a corpus of natural language documents into a structured ontology usable as such. Whereas the transformation from unstructured to semi-structured is a manual process, the transformation from semi-structured to structured is automatic, thus reducing the cost associated with changes in the semi-structured view.

The complete process from the corpus of text to the ontology is summarized in **¡Error! No se encuentra el origen de la referencia.**. This process (from left to right), describes the initial manual annotation in Obsidian, which allows both the XSD manual preparation / construction and the representation of the notes taken over the specifications using XML. Moreover, the notes can be automatically field from the initial annotations as described in 3.1. Once we have the XML/XSD information, the rest of the process is automatic. The ontology schema is created from the XSD and the XML notes are also automatically extended. Finally, with both the ontology schema and the notes, the ontology is automatically populated.

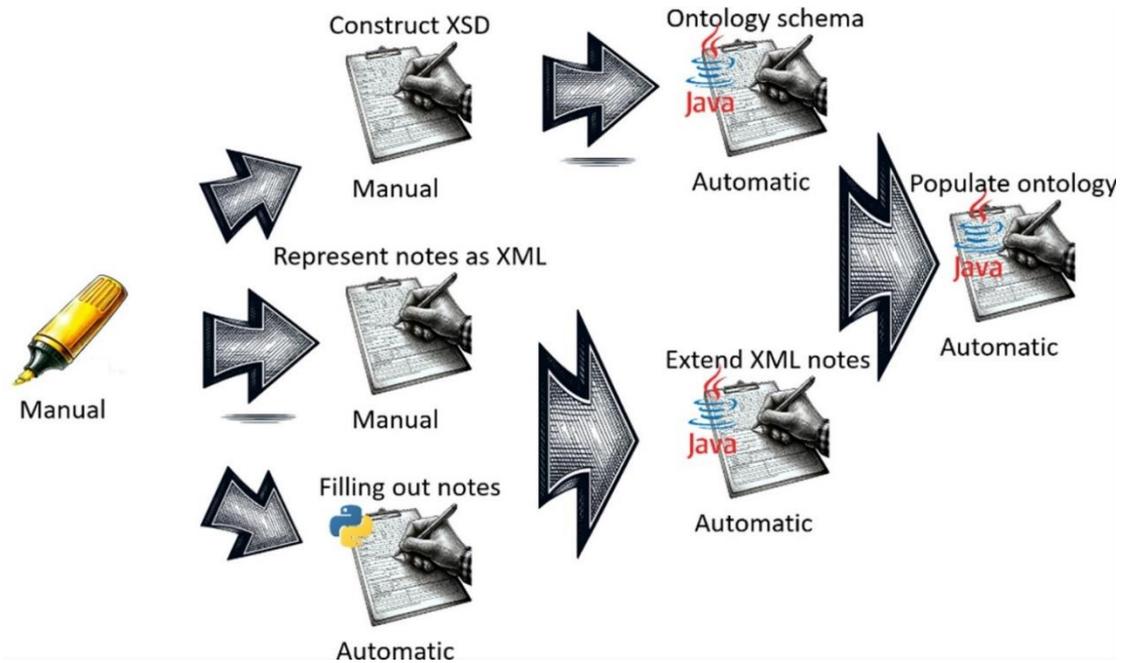

**Figure 4.** Ontology generation's software architecture

### 3.5. Node Fusing through LLM

As we have seen, the XML has been directly obtained from the specifications, therefore it contains a transcription of the specifications' requirements. When analyzing the ontology, we find that it is necessary to optimize its data because different names are often used to refer to identical or closely related concepts, resulting in a larger and unnecessary number of instances. For instance, in our context, "device", "medical device" and "IoT device" refer to the same thing.

As a way to address this, we want to detect redundant instances, to merge them into one. To be specific, we have used a Large Language Model (LLM), Llama [31], to do so. First, we send requests over the Llama API asking whether two instances might be merged, and require a True/False response (as shown in Figure 5). To do so, we use the default Llama model, i.e., a model without a specific training data pool. If Llama responds true, we add the tuple of the two instances to the short-list of nodes to merge.

We represent this short-list as a graph. In this structure, vertices represent ontology instances and edges indicate the possibility of merging. Based on these responses, we use graph theory to create a structure that identifies subsets of mergeable instances. With "subset of mergeable instances", we are referring to a clique, i.e., subsets of vertices such that there is an edge connecting every vertex to all others. To discover these cliques, we use the Bron-Kerbosch recursive algorithm [32].

The merging process consists in updating the domain and range of all properties of those instances that can be combined: all properties pointing from or to any of these instances will be redefined to point from or to the most representative instance. Through this process we do not lose information, yet we obtain a more robust and efficient structure while maintaining the integrity of the original data. Using this method, we have been able to reduce by 26% the number of instances.

To develop this solution, the OWL API [30] has been crucial, as it eases operations on ontologies expressed in OWL.

### 4. Results: case study prototype

The result of this process is an ontology, whose classes are described next. The ability to generate the ontology from the XML file simplifies the inclusion of new information. Furthermore, an example on how to query the ontology is provided.

```
1   {
2     "messages": [
3       {
4         "role": "system",
5         "content": "You only answer with true or false."
6       },
7       {
8         "role": "user",
9         "content": "Can I merge instances \"decommissioning procedure\" and \"issue addressing
            plan\"?"
10      }
11    ]
12  }
```

**Figure 5.** Request sent to Llama API

### 4.1. Ontology classes and subclasses

Clause 3.3 introduces some of the existing XML types and how we have applied them to our intermediate XML Schema, and which are thus included in the ontology. The choice which types are selected and included in XML Schema is dictated by the necessity of representing the specified requirements. Integrating a new specification might lead to the definition of new types and object properties.

The following is a non-exhaustive list of some of the types currently included in the ontology:

- Component
  - Storage component
- Content
  - Password
  - Record
  - Role
  - Skill
  - Threat
  - Constraint
  - Assurance
  - Decision
- Event
  - Change event
  - Failure event
  - Success event
- Requirement
  - Ensuring requirement
  - Design goal requirement
- Action
- Risk

As shown in clause 3.3 , we leverage the support of extensions to refine the definition of some of the concepts, for example to define more precisely the properties' range and domain. Some distinctions are harder to model through inheritance. For example, "risk" frequently appears with a subset of adjectives such as "(un)acceptable", "residual" or "identified". To avoid creating different types for each combination, we rather rely on the definition of children elements of Boolean type, which will be translated to data properties in the ontology. Figure 6 shows these properties for the case of the "risk" type.

```xml
<xs:complexType name="Risk">
  <xs:complexContent>
    <xs:extension base="ns:AbstractContent">
      <xs:sequence>
        <xs:element name="residual" type="xs:boolean" minOccurs="0"
            default="false"/>
        <xs:element name="accepted" type="xs:boolean" minOccurs="0"
            default="false"/>
        <xs:element name="unacceptable" type="xs:boolean" minOccurs="0"
            default="false"/>
        <xs:element name="identified" type="xs:boolean" minOccurs="0"
            default="false"/>
      </xs:sequence>
    </xs:extension>
  </xs:complexContent>
</xs:complexType>
```

**Figure 6.** Boolean elements used to describe the Risk type

### 4.2. Ease of adding new information

During the work carried out in MedSecurance project, new necessities have emerged: for example, being able to link identified vulnerabilities to requirement which are not complied with if said vulnerability is present. As our ontology relates the requirement to the specifications defining them, this also means being able to search, based on CVE (Common Vulnerabilities and Exposures) [33] [34], which specification compliance claims are not possible.

Although the CVEs are not part of the data originally considered for the ontology, thanks to the automatic translation from XML to ontology described in 3.4, we were able to easily adapt us to this new necessity.

We first modified the XSD to support the "invalidating CVEs" element for the base requirement type. Upon receiving the list of CVEs of interest to the project, we used ChatGPT to select the most relevant requirements, manually checked the selections and introduced them in the XML.

After a new execution of the transformation program, the ontology integrates the CVEs.

### 4.3. How to query the ontology: retrieving documentation requirements

The ontology is provided as a service through the definition of a Docker container executing a Jena-Fuseki [34][35] (an open-source SPARQL server) instance providing the data contained in the ontology through SPARQL.

During the building of the Docker container, the ontology is transformed in the model required for Jena-Fuseki's operation.

Figure 7 shows an example of query to retrieve the documentation requirements. Through the usage of the Optional syntax, the query adapts itself to the two main documentation requirement types: documenting actions and documenting general content.

As described in clause 3.3.1, within the XSD we rely on choices when multiple classes could be relevant. Therefore, the query shown in

Figure 7 must travel potentially with three different choices: the choice regarding the action' input's type, the action's output's type or the documented content's type.

```
1   PREFIX foaf: <http://xmlns.com/foaf/0.1/>
2   PREFIX rdf: <http://www.w3.org/1999/02/22-rdf-syntax-ns>
3   PREFIX rdfs: <http://www.w3.org/2000/01/rdf-schema>
4   SELECT ?documentedContentName ?documentedActionName ?documentedActionType
5          ?documentedActionInputContentName ?documentedActionOutputContentName {
6       #Selecting Documentation requirements, and retrieving their content
7       ?origin a foaf:DocumentationRequirement;
            foaf:DocumentationRequirement_choice ?content.
8       OPTIONAL {
9           #If the documentation requirement refers to an action, we retrieve the
                action input and output if aplicable
10          ?content a foaf:action; foaf:action_type ?documentedActionType;
                foaf:name ?documentedActionName.
11          OPTIONAL {
12              # If there is an input content, we retrieve its name
13              ?content foaf:action_choice ?actionChoice.
14              ?actionChoice foaf:Content_choice ?actionContentChoice.
15              ?actionContentChoice foaf:name ?documentedActionInputContentName.
16          }
17          OPTIONAL {
18              # If there is an output content, we retrieve its name
19              ?content foaf:output ?output.
20              ?output foaf:Content_choice ?outputContentChoice.
21              ?outputContentChoice foaf:name ?documentedActionOutputContentName.
22          }
23      }
24      OPTIONAL {
25          #If the document requirement refers to a general content, we retrieve
                its name
26          ?content a foaf:Content; foaf:Content_choice ?contentContentChoice.
27          ?contentContentChoice foaf:name ?documentedContentName.
28      }
```

**Figure 7.** Example SPARQL query to retrieve documentation requirements

## 5. Discussion

The presented approach rises some points of discussion: some points might be shared with other ontology constructions (available information which remains unused), other might be specific to this approach (due to the challenges of using an LLM, and the decisions which follow). We address these points next.

### 5.1. Unused specification content

Our approach relies on the formulation of the requirements, which implies that for now we do not integrate other, and also relevant, knowledge. For example, [13] states the requirement "Requirement that the operator should adopt cybersecurity good practices", which is then followed by a list of examples of such practices. This list includes for example integrity protection (where hashing is given as a sub-example) but does not include encryption (which is addressed in other clauses).

At the moment, our ontology does not establish the relationship between "cybersecurity good practice" and its example, and when addressing this shortcoming we will be faced with the decision to whether translate the explicit list of the clause, or rather what the specification seems to intend.

In specifications, frequently we encounter wording such as "symmetric encryption shall be used (e.g., AES)". This formulation has shortcomings when trying to establish the entire list of options. Whereas

AES [36] is given as an example, Twofish [37] is not. In other words, the specifications dedicated to IoMT do not appear to be a reliable source for a complete listing of valid options: the requirement is defined on selecting and using a solution compliant with the currently accepted understanding of the problem. Linking with existing ontologies, such as [8], would allow to mitigate this issue.

In order to ensure the completeness of our examples, it might be preferable to link the concepts relevant to the requirements to the instances of existing ontologies, for example regarding the attacks, which are included in the analyzed ontologies (see Table 1).

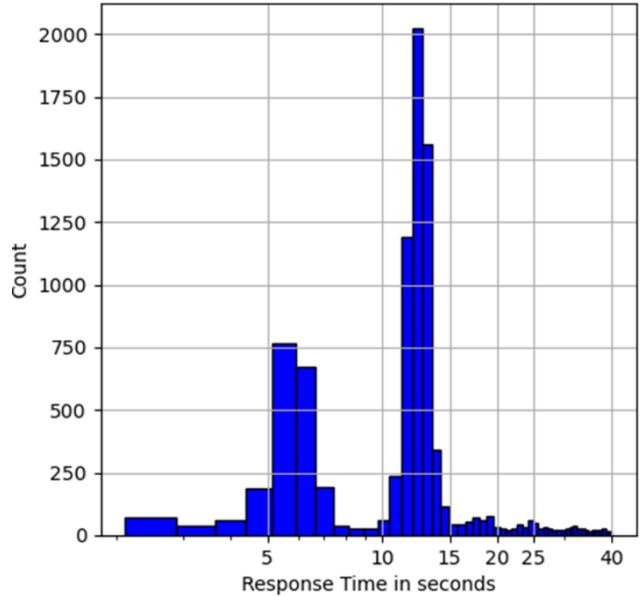

**Figure 8.** Histogram of Llama queries' times

### 5.2. Challenges when using LLM

Despite the potential of the Llama API, its reliability is not absolute, and the lack of context may lead Llama to provide incorrect responses. Consequently, a human review is required to ensure the correctness of the performed merges. To evaluate this performance, we have obtained several measures. Our results show a precision of 25.96% and a recall of 92.5%, leading to an F-score of approximately 40.55%. This indicates that, while the API is effective at identifying instances that should not be merged, it tends to label instances as mergeable too frequently (i.e., a high number of false positives).

Regarding the evaluation of response times, Figure 8 shows the distribution of times. The responses are mainly concentrated between 2.5 and 20 seconds, with a mean response time of approximately 14 seconds from the moment the API call was made until a response was received. This implies that it is not feasible to test the mergeability of all instances against all instances, at least with our current prompt.

We started using Llama API with an initial voucher of $5 [38]. After 20680 calls, we have consumed $0.724. The low cost per request suggests that the Llama API is a viable option for applications that require iterative interactions, provided that the issue of false positives can be mitigated.

In conclusion, while the Llama API demonstrates potential and cost efficiency in refining the generated ontology, its high rate of false positives requires human judgment to determine the appropriate action in the merge process to maintain data integrity and accuracy.

Possible solution could include training the model on our corpus, but this would require more computational resources, or find a different prompt where false positives are less likely. Since the refinement is non-destructive, the results of the LLM can be checked by the user and does not affect negatively the generated ontology.

We have also tested a different approach on ChatGPT, by leveraging the fact that we were able to upload a file. Instead of asking the LLM about each pair, we provided the list of instance names, and prompted to have them grouped in mergeable sets. Whereas Llama was more prone to cause false-positives, ChatGPT's responses have more false-negatives: while the precision improves to 76.3%, the recall descends to 19.7%, leading to a lower F-score of 31.25%. When trying to obtain the same results a second time, ChatGPT showed a high variability in the results (precision: 36.76%, recall 52.5% and F-score 43.24%).

Since we are requesting from ChatGPT a grouping of entries instead of the mergeability of two given entries, we try to adapt our ground-truth to this use case by using the result of the clique computation. This does improve the recall for the two test cases (from 10.7 to 21.6% and 52.5 to 86.1%), but the precision is less good leading to a lower F-score (20% and 34.6%, respectively).

The obtained results are not convincing enough to use the responses from the LLMs without supervision. As previously mentioned, further training the model on our corpus, might improve the results. Another solution might be to retrieve from the Obsidian notes all the natural language text related to the topics at hand, and provide this information to the LLM alongside our question.

**5.3. Refinement decisions**

As explained in clause 3.5, we aim at improving the legibility of the ontology through its simplification. We intend to achieve this simplification by merging similar instances. However, the decision of merging two nodes might be debated. For example, [14] includes "Requirement that the manufacturer shall encrypt and authenticate data exchange", and [11] includes "Requirement that the manufacturer shall make the system to use data encryption at rest and in transit". One could argue that both requirements are similar enough and should be merged.

However, the encryption of data at rest might have implications on safety, for example in a case where "break- the-glass" would be legitimate. From this point of view, merging these requirements is less clear decision. Thus, different actors might have different opinions on the correct approach on a case-by-case basis: to mitigate this issue, our approach is non-destructive and allows to review the previous decisions later on.

# 6. Conclusion

To facilitate the creation of tools guiding a user through the development of a secure IoMT device or service, we structured the information contained in relevant security and safety specifications in an ontology. The objective is to simplify the retrieval of requirements and their related concepts.

As the information contained in the specifications is in an unstructured natural language form, we have used multiple steps to structure the data. The first step has been highlighting relevant portions of the text using Obsidian MD [2]. We then manually semi-structured this summarized information in an XML file (alongside its XSD schema file we defined ourselves). The final step to obtain the ontology is to use the tools we have developed to transform this XML/XSD into the OWL-RDF ontology. This tool places some constraints both on the XML/XSD and the subsequent queries to the ontology.

We have used our pipeline to create an ontology dedicated to the requirements and concepts based on the specifications (International standards) identified as relevant to the field. The fact that we have a tool dedicated to transforming the XML into the ontology, also simplifies the addition of new information entries to existing elements. We used this feature to insert the Common Vulnerabilities and Exposures (CVEs) relevant to a given requirement.

The different specifications might refer to the same or similar concepts, with different names. This introduces complexity in the ontology by adding unnecessary instances. We query an LLM to obtain a short-list of instances which could be merged. After a manual check of this list, we decide whether to perform or not the merging of the selected nodes. Besides the challenge to process all instances to construct the candidates short-list, there is also a comprehension challenge. In some instances, the differences in wording might reflect real differences which should not be overlooked. The manual check therefore should not be limited to the instance name (the information currently provided to the LLM), but should also include the background information.

In this future work task, the presented ontology should be extended with information from other ontologies (for example the ones introduced in subclause 2.1). This will allow to cover a broader range of concepts, as for example the current proposal lacks in terms of documented attacks.

# Acknowledgements

The work presented in this paper has been partially supported by the Spanish Government under the project GenClinLab-Sec (PID2020-114394RB-C31) funded by MCIN/AEI/10.13039/501100011033, by the Generalitat de Catalunya (2021 SGR 01252) and by the HORIZON.2.1 - Health Programme of the European Commission, Grant Agreement number: 101095448 - MEDSECURANCE.